\documentclass[12pt]{JHEP3}
\usepackage{amsmath,amsfonts,amssymb}
\usepackage{verbatim,graphicx,float}

\title{Effective dynamics of the closed \\ loop quantum cosmology}

\author{Jakub Mielczarek$^a$\footnote{jakub.mielczarek@uj.edu.pl}, 
Orest Hrycyna$^{b}$\footnote{hrycyna@kul.lublin.pl}
and Marek Szyd{\l}owski$^{abc}$\footnote{uoszydlo@cyf-kr.edu.pl}  \\
$^a$ Astronomical Observatory, Jagiellonian University,\\ 30-244
Krak\'ow, Orla 171, Poland \\
$^b$ Department of Theoretical Physics, 
Catholic University of Lublin,\\
 Al. Rac{\l}awickie 14, 20-950 Lublin, Poland \\
$^c$ Marc Kac Complex Systems Research Centre,\\ Jagiellonian University,
Reymonta 4, 30-059 Krak{\'o}w, Poland 
}

\abstract{
In this paper we study dynamics of the closed FRW model with holonomy corrections 
coming from loop quantum cosmology. We consider models with a scalar field and 
cosmological constant. In case of the models with cosmological constant and free scalar 
field, dynamics reduce to 2D system and analysis of solutions simplify. If only 
free scalar field is included then universe undergoes non-singular oscillations. For the 
model with cosmological constant, different behaviours are obtained depending on the  
value of $\Lambda$. If the value of $\Lambda$ is sufficiently small, bouncing solutions 
with asymptotic de Sitter stages are obtained. However if the value of $\Lambda$ exceeds
critical value $\Lambda_{\text{c}} = \frac{\sqrt{3} m^2_{\text{Pl}}}{2\pi\gamma^3} 
\simeq 21 m^2_{\text{Pl}}$ then solutions become oscillatory. Subsequently we study 
models with a massive scalar field. We find that this model possess generic inflationary 
attractors. In particular field, initially situated in the bottom of the potential, is 
driven up during the phase of quantum bounce. This subsequently leads to the phase 
of inflation. Finally we find that, comparing with the flat case, effects of curvature 
do not change qualitatively dynamics close to the phase of bounce. Possible effects of 
inverse volume corrections are also briefly discussed. 
}

\begin{document}

\section{Introduction} \label{Intro}

In the recent years  methods of the Loop Quantum Gravity (LQG) \cite{Ashtekar:2004eh} 
have been successfully applied to quantise cosmological minisuperspace models. General 
prediction of this approach, called Loop Quantum Cosmology (LQC) \cite{Bojowald:2008zzb}, 
is avoidance of the initial singularity. Namely in LQC singularity is replaced by the phase of 
bounce. This result seems to be generic for the homogeneous cosmologies and was confirmed 
in the numerous investigations. Loop quantisation of the homogeneous and isotropic FRW 
models have been performed in Ref. \cite{Ashtekar:2006rx,Ashtekar:2006uz,Ashtekar:2006wn} ($K=0$), 
Ref. \cite{Ashtekar:2006es,Szulc:2006ep} ($K=1$) and Ref. \cite{Vandersloot:2006ws,Szulc:2007uk} ($K=-1$). 
In the papers cited above strict, fully quantum considerations were performed. However 
models only with a free scalar field and cosmological constant were studied. 

Dynamics of the models also can be traced with the effective equations of motion. Then classical dynamics 
is modified with certain nonperturbative quantum gravitational corrections. These corrections 
becomes unimportant for the low energy densities leading to the classical equations of motion.
However for the scales comparable with the Planck energy densities corrections produce effective  
``quantum repulsion'' leading to the bounce. Studies of the effective dynamics have recently attracted
substantial interest. It is mainly due to the fact that it gives the access to studying various physical 
consequences of the loop quantisation. Moreover, semi-classical considerations are in many points 
simpler than the fully quantum treatment. Often approximate solution can be found before
the strict quantum results are obtained. In particular models with self-interacting scalar field
can be easily studied in this effective formulation. Therefore both fully quantum and effective methods 
are complementary to each other. Since now effective flat FRW \cite{Singh:2006im,Mielczarek:2008zv} 
and Bianchi I \cite{Chiou:2007mg} models in LQC were studied. In the present paper we study effective 
dynamics of the closed ($K=1$) FRW model.

We have to stress that alternative effective description is developed by Bojowald and his collaborators.
In the Bojowald's approach information about dispersions and higher moments of the observables 
can also be obtained. Recently this approach has been applied to Bianchi I model \cite{Chiou:2008bw}. 
However we do not apply this approach in our considerations. Recently also new approach to quantise 
loop quantum cosmological models has been proposed \cite{Malkiewicz:2009zd,Dzierzak:2008dy}. This formulation
leads to the bouncing solutions but no effects of so called Thiemann's term, which leads to inverse 
volume corrections, are present. Obtained effective dynamics is therefore the same like this studied in 
\cite{Singh:2006im,Mielczarek:2008zv}  where only holonomy corrections were taken into account. 

Canonical variables for the closed FRW model are similar like in the flat case. Namely original 
Ashtekar variables $(A^i_a,E^a_i)$ are parametrised by the conjugated $(c,p)$ variables with 
the Poisson bracket 
\begin{equation}
\left\{ c,p \right\} = \frac{8\pi G \gamma}{3}. 
\end{equation}
Parameters $(c,p)$ can be expressed in terms of the standard FRW variables in the 
following way
\begin{eqnarray}
p = a^2  \left( \frac{l_0}{a_0} \right)^2 \ \ \text{and} \ \ 
c = \gamma \dot{a}  \frac{l_0}{a_0}
\end{eqnarray}
where $a_0=2$, fiducial volume $V_0 = 2\pi^2 a_0^3 = 16 \pi^2$ and $l_0 = V_0^{1/3} = 
(16 \pi^2 )^{1/3} = (2 \pi^2 )^{1/3} a_0$. It will become useful when correspondence with
classical case will be studied. 

In this paper value of Barbero-Immirzi parameter $\gamma = 0.2375$, following  Ref. \cite{Meissner:2004ju}, is assumed.
We also use notation $\kappa =8\pi G$.

\section{Effective equations for the closed FRW cosmology}

Starting point of our considerations is the effective Hamiltonian 
\begin{equation}
\mathcal{H}_{\text{eff}} = \frac{-3\sqrt{p} N }{8 \pi G \gamma^2 \bar{\mu}^2} 
\left[  \sin^2\left(\bar{\mu} c - \bar{\mu}\frac{l_0}{2}  \right)   
- \sin^2 \left( \bar{\mu}\frac{l_0}{2}  \right)   +(1+\gamma^2) \frac{\bar{\mu}^2 l_0^2}{4}  \right]
+ \mathcal{H}_{\text{m}}   \label{Hamiltonian1} 
\end{equation}
which was derived in Ref. \cite{Ashtekar:2006es}. Here $\bar{\mu}$ is quantisation parameter 
and its form depends on the particular quantisation scheme. In this paper we choose so 
called $\bar{\mu}$-scheme, then    
\begin{eqnarray}
\bar{\mu}=\sqrt{\frac{\Delta}{p}}
\end{eqnarray}  
where $\Delta = 2\sqrt{3} \pi \gamma l_{\text{Pl}}^2$. Earlier, $\mu_0$-scheme was also used, 
however it was realised later that it has wrong classical limit.   
Recently other quantisation scheme was proposed \cite{Mielczarek:2008zz} and studied in 
Ref. \cite{Hrycyna:2008yu,Li:2008tg}. In this scheme new types of finite scale factor 
singularities appears. Another possibility is those presented in \cite{Malkiewicz:2009zd,Dzierzak:2008dy}
where $\bar{\mu}=\lambda/\sqrt{p}$ and $\lambda $ is some unknown constant unrelated with $\Delta$.

As a matter content we consider homogeneous scalar field with the Hamiltonian 
\begin{equation}
\mathcal{H}_{\text{m}} = Np^{3/2}\left(\frac{p^2_{\phi}}{2 p^{3}}+V(\phi)\right).   
\end{equation}
With this assumption it will be possible to study realisation of the inflationary phase
in the bouncing universe. Energy density of the scalar field can be derived from 
the following expression
\begin{equation}
\rho := \frac{1}{p^{3/2}} \frac{\partial \mathcal{H}_{\text{m}}}{\partial N}.
\end{equation}

Since the Hamiltonian constraint $\frac{\partial \mathcal{H}_{\text{eff}}}{\partial N} = 0 $
is fulfilled we obtain
\begin{equation}
\sin^2\left(\bar{\mu} c - \bar{\mu}\frac{l_0}{2}  \right)  =  \left[   \sin^2 \left( \bar{\mu}\frac{l_0}{2}  \right) 
 -(1+\gamma^2) \frac{\bar{\mu}^2 l_0^2}{4}    \right] +\frac{\rho}{\rho_c} \label{HamConstr}
\end{equation}
where we have defined critical density
\begin{equation}
\rho_{\text{c}} = \frac{3}{\kappa \gamma^2 \Delta} =\frac{\sqrt{3}}{16\pi^2\gamma^3 l^4_{\text{Pl}} }
\simeq 0.82 \rho_{\text{Pl}}.
\end{equation}
In the rest of the paper we choose the gauge $N=1$.

The equations of motion can be derived with the use of Hamilton
equation 
\begin{equation}
\dot{f} =  \{f, \mathcal{H}_{\text{eff}}  \} \label{HamiltonEquation}
\end{equation}
where the Poisson bracket is defined as follows
\begin{equation}
\{f,g\} = \frac{8 \pi G \gamma }{3}
\left[\frac{\partial f}{\partial c}\frac{\partial g}{\partial p}- 
\frac{\partial f}{\partial p}\frac{\partial g}{\partial c}  \right]   
+ \left[\frac{\partial f}{\partial \phi}\frac{\partial g}{\partial p_{\phi} }-
\frac{\partial f}{\partial p_{\phi}}\frac{\partial g}{\partial \phi}  \right].
\end{equation}
From this definition we can retrieve the elementary brackets
\begin{equation}
\{c,p \} = \frac{8 \pi G \gamma }{3} \ \ \text{and} \ \  \{\phi,p_{\phi} \} = 1 .
\end{equation}
Based on Eq. \ref{HamiltonEquation} we derive equation for the scalar field part 
\begin{eqnarray}
\dot{\phi} &=& \{\phi,\mathcal{H}_{\text{eff}}  \}=  p^{-3/2}  p_{\phi}, \\
\dot{p_\phi} &=& \{p_\phi, \mathcal{H}_{\text{eff}}  \}=-p^{3/2}\frac{dV}{d\phi} 
\end{eqnarray}
what leads to the classical equation
\begin{equation}
\ddot{\phi}+\frac{3}{2} \frac{\dot{p}}{p} \dot{\phi}+\frac{dV}{d\phi} = 0. \label{ScalarEquation}
\end{equation}
Here equation of motion for the scalar fields does not feel the quantum corrections.

For the gravity part we calculate
\begin{equation}
\dot{p} =  \{p, \mathcal{H}_{\text{eff}}  \} =- \frac{8 \pi G \gamma }{3}
 \frac{\partial \mathcal{H}_{\text{eff}} }{\partial c}
\end{equation}
what together with Eq. \ref{HamConstr} give us effective Friedmann equation
\begin{equation}
H^2 := \left( \frac{\dot{p}}{2p} \right)^2=
\frac{8\pi G}{3}  \frac{1}{ \rho_{\text{c}} } (\rho-\rho_1(p))(\rho_2(p)-\rho) \label{ModFried}
\end{equation}
where
\begin{eqnarray}
\rho_1(p) &=& -  \rho_{\text{c}}  \left[ 
    \sin^2 \left( \sqrt{\frac{\Delta}{p}}   \frac{l_0}{2}  \right) 
 -(1+\gamma^2) \frac{\Delta l_0^2}{4 p} \right]  \approx  \frac{3}{\kappa a^2},  \label{rho1}   \\
\rho_2(p) &=&  \rho_{\text{c}} 
\left[ 1-\sin^2 \left( \sqrt{\frac{\Delta}{p}} \frac{l_0}{2}  \right)+(1+\gamma^2) \frac{\Delta l_0^2}{4 p}   \right]
\approx \rho_{\text{c}}+\frac{3}{\kappa a^2}. \label{rho2} 
\end{eqnarray}
It is clear from the above expressions that relation $\rho_2(p)-\rho_1(p) = \rho_{\text{c}}$
is fulfilled. Approximations performed in (\ref{rho1}) and (\ref{rho2}) are valid for
$p \gg  \frac{\Delta l_0^2}{4} \simeq 18.9 \ l^2_{\text{Pl}}$.

\begin{figure}[ht!]
\centering
\includegraphics[width=8cm,angle=0]{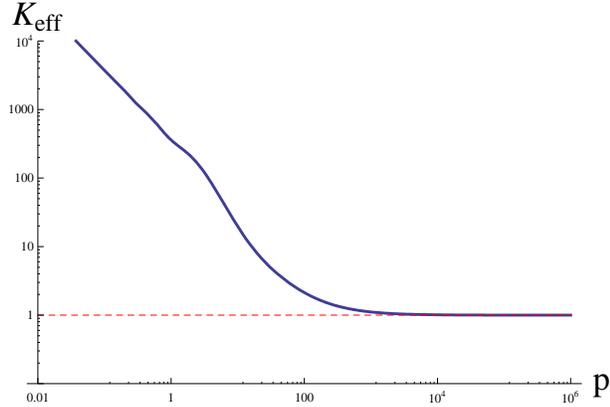}
\caption{Effective parameter of the curvature $K_{\text{eff}}$.}
\label{Keffpict}
\end{figure}

Alternatively expression for the modified Friedmann equation can be written as follows
\begin{equation}
H^2 = \frac{8\pi G_{\text{eff}}}{3}  \rho -\frac{K_{\text{eff}}}{a^2}  
\end{equation}
where effective constants are expressed as follows
\begin{eqnarray}
  G_{\text{eff}} &=& G \left[ 1+2\frac{\rho_1}{\rho_{\text{c}}}-\frac{\rho}{\rho_{\text{c}}} \right], \\
K_{\text{eff}} &=&  \frac{\kappa}{3} \left( \frac{a_0}{l_0}\right)^2 p \frac{\rho_1(p)\rho_2(p)}{\rho_{\text{c}}}.  
\end{eqnarray}
Expression for the $K_{\text{eff}}$ does not depends on the energy density $\rho$. Therefore 
it can be studied without specifying form of the matter. We expect that classical limit
$K_{\text{eff}} \rightarrow 1$ for $p \rightarrow \infty$ should be recovered.
It is in fact a case, what it is clear from Fig. \ref{Keffpict}. 
For $p \rightarrow 0$, factor $K_{\text{eff}}$ grows and can potentially play important role.
We study this issue in the subsequent sections.

Defining energy density and pressure of the scalar field 
\begin{eqnarray}
\rho         &=& \frac{\dot{\phi}^2}{2}+V(\phi), \\
\mathcal{P} &=& \frac{\dot{\phi}^2}{2}-V(\phi).
\end{eqnarray} 
we can derive equation
\begin{eqnarray}
\frac{dH}{dt} = -4\pi G \left[(\rho +\mathcal{P}) - M(p)\rho_{\text{c}} \right] 
\left[1+2\frac{\rho_1}{\rho_{\text{c}}}-2\frac{\rho}{\rho_{\text{c}}} \right] \label{RauMod}
\end{eqnarray}
where to simplify notation we have defined
\begin{eqnarray}
M(p) =  (1+\gamma^2) \frac{\Delta l_0^2}{6 p} - \frac{l_0}{3} \sqrt{\frac{\Delta}{p}}
\sin \left( \sqrt{\frac{\Delta}{p}}\frac{l_0}{2}  \right)\cos \left( \sqrt{\frac{\Delta}{p}}\frac{l_0}{2}  \right).
\end{eqnarray}

It is worth to mention that taking $l_0=0$ in (\ref{ModFried}) and (\ref{RauMod}) we obtain
\begin{eqnarray}
H^2 &=& \frac{8\pi G}{3}  \rho \left(1 - \frac{\rho}{\rho_{\text{c}} }\right), \\
\frac{dH}{dt} &=& -4 \pi G (\rho +\mathcal{P}) \left[1 - 2 \frac{\rho}{\rho_{\text{c}}} \right]
\end{eqnarray}
recovering the case $K=0$.

\section{Models with $\Lambda$ and free scalar field as 2D dynamical systems}

Effective loop dynamics of the flat FRW models with $\Lambda$ and free scalar field
was studied in \cite{Mielczarek:2008zv}. In this reference, fully analytical solutions 
of the model were found. Unfortunately in case of the closed model the equations become
transcendental and cannot be solved analytically. Therefore other approach, to investigate dynamics,
have to be applied. In our considerations we use methods of the qualitative analysis of 
the dynamics. Useful technique applied here is compactification of the phase space. Based 
on this approach it will be possible to study global properties of dynamics. 

In this section we consider models with cosmological constant and free scalar field. For these 
models dynamics reduce to 2D dynamical systems, what simplify qualitative analysis. Moreover one 
has to remember that physical solutions are on the surface of Hamiltonian constraint 
$\mathcal{H}_{\text{eff}}(p,\dot{p})\approx 0$. Therefore domain allowed for the motion reduce 
to one dimensional subspace. It means that on 2D phase plane $(p,\dot{p})$, only one trajectory represents 
physical motion. Other possible trajectories does not fulfil $\mathcal{H}_{\text{eff}}(p,\dot{p})\approx 0$.
It is  also interesting to note that in all cases, the same maximal value of Hubble parameter 
$H_{\text{max}}$ is reached. In fact this is the same $H_{\text{max}}=\sqrt{\frac{\kappa}{12} \rho_c}$ 
as this obtained for the flat models. 

\subsection{Cosmological constant}

In the first case we consider model with cosmological constant only.
Energy density and pressure of the cosmological constant are given by 
\begin{eqnarray}
\rho_{\Lambda} = \frac{\Lambda}{8\pi G} \ \ \text{and} \ \ 
\mathcal{P}_{\Lambda} = - \frac{\Lambda}{8\pi G}.
\end{eqnarray} 

\begin{figure}[ht!]
\centering
\includegraphics[width=8cm,angle=0]{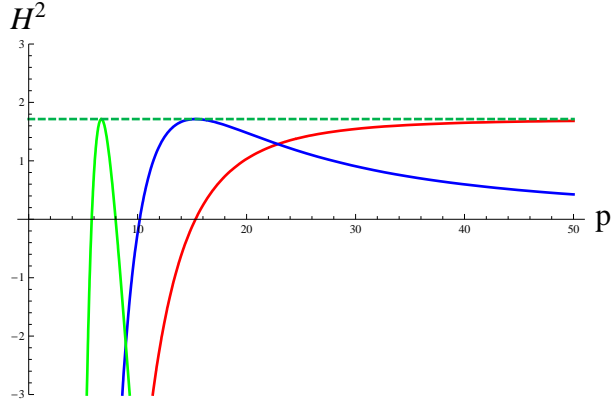}
\caption{Square of the Hubble factor for different values of $\rho_{\Lambda}$.
From right $\rho_{\Lambda}=0.5 \rho_c$ (red), $\rho_{\Lambda}= \rho_c$ (blue) and 
$\rho_{\Lambda}=2.5 \rho_c$ (green). Dashed line represents $H^2_{\text{max}}=\frac{\kappa}{12} \rho_c $.}
\label{H2Lambda}
\end{figure}

Dynamics of the system is governed now only by modified Friedmann equation (\ref{ModFried}).
Since left side of this equation is $H^2\geq 0$ on can find physical domain of motion
form this condition. We plot $H^2$ function given by Eq. (\ref{ModFried}) in Fig. \ref{H2Lambda}.
We find that while $\rho_{\Lambda}<\rho_c$ then domain for the physical motion
is bounded from two sides. Therefore we expect oscillatory solutions in this regime. 
While $\rho_{\Lambda}=\rho_c$ the upper bound reach infinity. In turn, when $\rho_{\Lambda}>\rho_c$
upper bound disappears. It is useful to define critical value of cosmological constant 
\begin{equation}
\Lambda_{\text{c}} = \frac{\sqrt{3} m^2_{\text{Pl}}}{2\pi\gamma^3} 
\simeq 21 m^2_{\text{Pl}}
\end{equation}
since bifurcation of solutions occur when value of cosmological constant cross $\Lambda_{\text{c}}$.
This critical value was introduced in Ref. \cite{Mielczarek:2008zv} while studying flat models in LQC.
It was show there that there is no physical solutions for $\Lambda>\Lambda_{\text{c}}$. In the curved 
models we however expect physical motion in this region. 

\begin{figure}[ht!]
\centering
a)\includegraphics[scale=1]{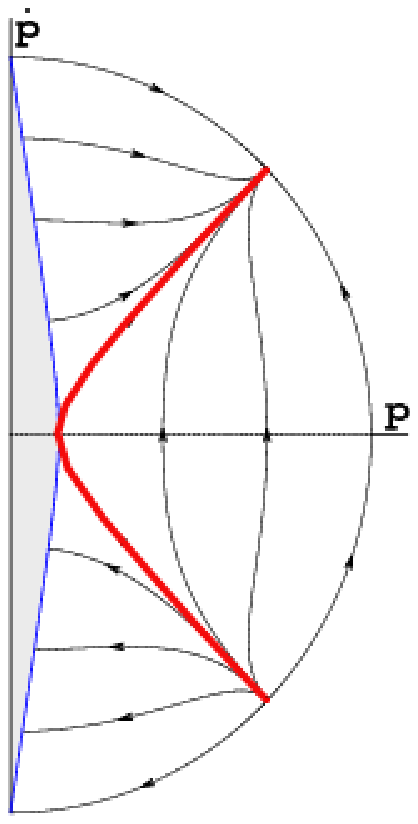}
b)\includegraphics[scale=1]{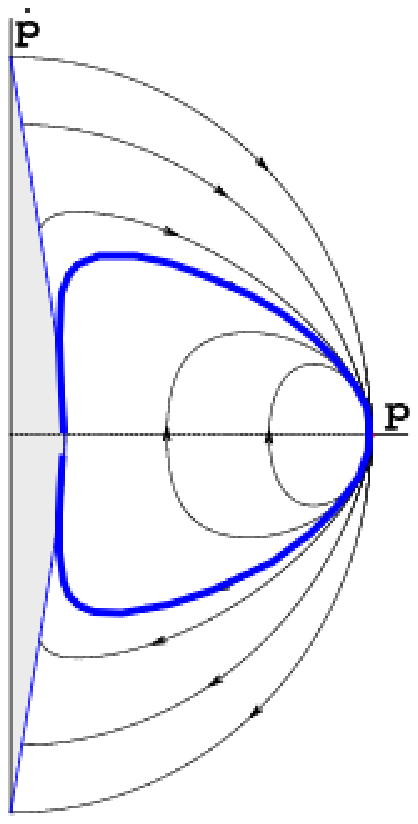}
c)\includegraphics[scale=1]{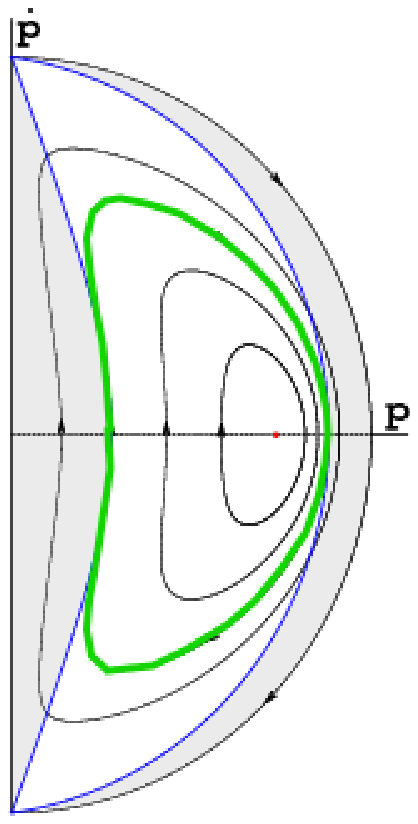}
\caption{Phase space diagrams for the model with the cosmological constant.
Shaded regions are unphysical because of $H^{2}<0$. Case a) 
$0<\rho_{\Lambda}< \rho_c$, b) $\rho_{\Lambda}=\rho_c$, c) $\rho_{\Lambda}>\rho_c$. 
Physical solutions, on the surface of Hamiltonian
constraint $\mathcal{H}_{\text{eff}}(p,\dot{p})\approx 0$, are distinguished by thick 
(red, blue, green) lines. }
\label{PPLambda}
\end{figure}

Now we can investigate dynamics of these three cases on the phase phase portraits shown in Fig. \ref{PPLambda}.
We see tree mentioned types of solutions. 
It is worth to stress that only thick lines represent physical trajectories. Other trajectories 
shown in Fig. \ref{PPLambda} are not on the surface of Hamiltonian constraint 
$\mathcal{H}_{\text{eff}}(p,\dot{p})\approx 0$. Based on approximations performed in 
(\ref{rho1}) and (\ref{rho2}) we find 
\begin{eqnarray}
p_{\text{min}} &\approx& \frac{3}{4} \frac{l^2_0}{\Lambda} \ \mapsto  a_{\text{min}} \approx \sqrt{\frac{3}{\Lambda}}, \\
p_{\text{max}} &\approx& \frac{3}{4} \frac{l^2_0}{\Lambda-\Lambda_{\text{c}}} \ \mapsto  a_{\text{max}} \approx \sqrt{\frac{3}{\Lambda-\Lambda_{\text{c}}}}.
\end{eqnarray}
Expression for $p_{\text{max}}$ is however valid only for $\Lambda \geq \Lambda_{\text{c}}$. 
When $\Lambda < \Lambda_{\text{c}}$ we have  $p_{\text{max}} \rightarrow \infty$. In this 
case asymptotic de Sitter stage is obtained as in the flat case.

\subsection{Free scalar field}

Now we get on to example of free scalar filed which is widely used in LQC. It is due 
to the fact that, in this model, scalar field can be treated as internal clock.  It is 
of great importance when purely quantum cosmological models are constructed. However it
is not necessary on the semi-classical level. Therefore models with no well defined 
global internal time can also be studied. Example of this will be given in the subsequent
section when model with self-interacting field is studied. However even in that 
case well defined intrinsic time can be introduced at intervals. 

Considered model with free scalar filed was studied before in Ref. \cite{Ashtekar:2006es}.
In this reference fully quantum analysis of the model has been done. In particular wave 
function for this model was computed. It was also shown that semi-classical dynamics
follows precisely the quantum evolution. It is important result approving applicability of the 
semi-classical approach. Here we would like to complete earlier investigations.

\begin{figure}[ht!]
\centering
\includegraphics[width=8cm,angle=0]{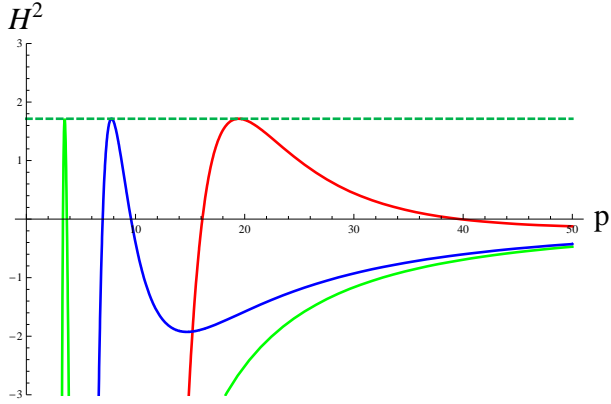}
\caption{Square of the Hubble factor for different values of $p_{\phi}$. From left to 
right $p_{\phi}=20, 40, 100 \ l_{\text{Pl}}$. Dashed line represents 
$H^2_{\text{max}}=\frac{\kappa}{12} \rho_c $.}
\label{H2FreeField}
\end{figure}

It is worth to remind here that energy density and pressure of the free scalar field are given by 
\begin{eqnarray}
\rho_{\phi}        = \frac{p^2_{\phi}}{2 p^{3}} \ \ \text{and} \ \
\mathcal{P}_{\phi} = \frac{p^2_{\phi}}{2 p^{3}}
\end{eqnarray}
where $p_{\phi}$ is constant of integration here. Based, as earlier, on Eq. (\ref{ModFried}) we can 
find the regions $H^2\geq 0$. This gives us regions allowed for motion. We show exemplary curves in 
Fig. \ref{H2FreeField}. It can be seen that doubly restrict region is allowed for the motion. 
This region is enlarged with increasing value of parameter $p_{\phi}$.

\begin{figure}[ht!]
\centering
\includegraphics[scale=1]{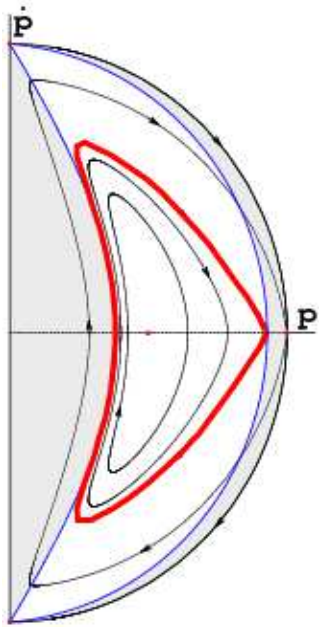}
\caption{Phase space diagram for the model with free scalar field. The shaded regions
are unphysical because $H^{2}<0$ there. Physical solution, on the surface of Hamiltonian
constraint $\mathcal{H}_{\text{eff}}(p,\dot{p})\approx 0$, is distinguished by thick (red) line.}
\label{PPFreeField}
\end{figure}

Based on analysis of positivity of $H^2$ function one expect that solutions are 
represented by oscillations. Namely universe bouncing between two edges of the region
allowed for motion. It can be approved looking on phase portrait for this system, shown 
in Fig. \ref{PPFreeField}. 
In fact physical solution is represented by closed curve. Similarly like in the previous case, 
physical solution lie on the surface of Hamiltonian constraint. Here we can also find 
\begin{eqnarray}
p_{\text{min}} \approx  \sqrt[3]{\frac{p^2_{\phi}}{2\rho_{\text{c}} }}  \ \ \text{and} \ \  
p_{\text{max}} \approx \sqrt{ \frac{\kappa}{3} \frac{2}{l^2_0} p^2_{\phi}}
\end{eqnarray}
based on approximations performed in (\ref{rho1}) and (\ref{rho2}). These approximations 
are valid for $p \gg  \frac{\Delta l_0^2}{4} \simeq 18.9 \ l^2_{\text{Pl}}$.

\section{Qualitative analysis of the model with quadratic potential function}

Self-interacting scalar fields are commonly used in the modern cosmology. It comes 
from the fact that violation of the strong energy condition $\rho+3p\geq 0$ can be 
obtained with them. Namely $\rho_{\phi}+3p_{\phi}=2(\dot{\phi}^2-V(\phi))$ and 
it is possible to have $V(\phi)>\dot{\phi}^2$ what leads to $\rho+3p<0$. In consequence 
accelerating stage of the evolution of the universe occurs. This mechanism can be 
therefore applied to modelling dynamics of the inflation or present era of dark energy.
Here we concentrate on this former. It is due to the fact that stage of inflation 
is proceeded by the Planck era and both phases should be dynamically related. Namely 
initial conditions for the inflation should be given in the Planck epoch. In our approach 
the Planck epoch is described by the closed loop cosmology. It is interesting to see how 
phase of inflation emerge in this scenario. In fact this was already studied in case
of the flat model \cite{Singh:2006im}. It was shown that phase of bounce can be naturally 
connected with the inflation. Moreover inflationary attractor is generic and is realised 
for the broad range of the initial conditions. Here we are going to check whether curvature 
can affect this scenario.

\begin{figure}[ht!]
\centering
\includegraphics[scale=1]{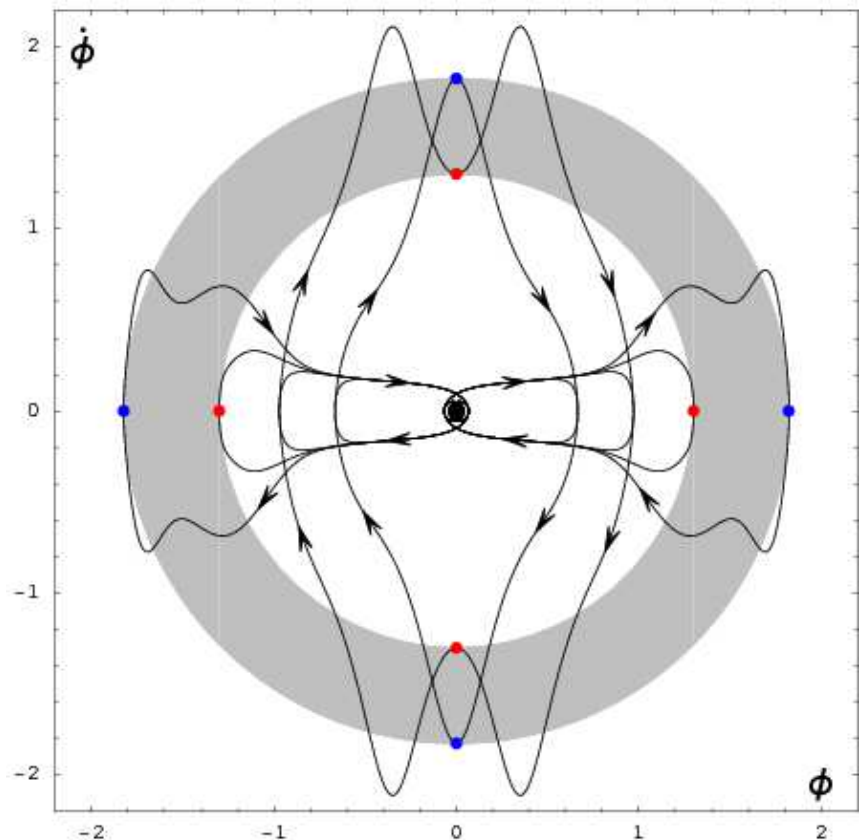}
\caption{Projection of the $3D$ phase space diagram for the model with the
quadratic potential function and $m^{2}=1$ on the $(\phi,\dot{\phi})$
plane. The shaded field denotes region admissible for motion for constant value
of $p=10$, size of this region depends on the value of $p$. The red and blue
dots represent location of the initial conditions taken for numerical
simulations.}
\label{phiphidot}
\end{figure}

\begin{figure}[ht!]
\centering
\includegraphics[scale=1]{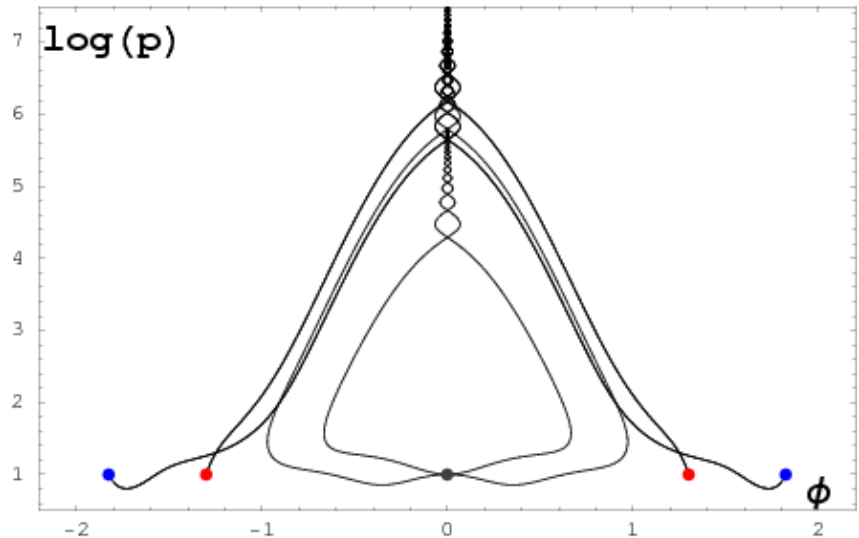}
\caption{Projection of the $3D$ phase space diagram for the model with the
quadratic potential function and $m^{2}=1$ on the $(\phi,\log{(p)})$
plane. The arrows on the trajectories are omitted due to symmetries (every line
can represent both expanding or contracting phase of evolution).}
\label{pphi}
\end{figure}

In our considerations we assume that considered scalar field is massive. Then energy density 
is given by  
\begin{equation}
\rho_{\phi}=\frac{1}{2}\dot{\phi}^{2}+V(\phi) = \frac{1}{2}\dot{\phi}^{2}+\frac{1}{2}m^{2}\phi^{2}.
\end{equation}
Dynamical system, in case of self-interacting scalar fields, can be written in the form  
\begin{eqnarray}
\dot{\phi} &=& y, \\
\dot{y}    &=& -\frac{3}{2}\frac{z}{p}y - \frac{dV(\phi)}{d\phi}, \\
\dot{p}    &=& z ,\\
\dot{z}    &=& 2\frac{\kappa}{3}\rho_{c} p \Bigg\{ 
2 \Big(\sin^{2}\big(\sqrt{\frac{\Delta}{p}}\frac{l_{0}}{2}\big) -
(1+\gamma^{2})\frac{\Delta l_{0}^{2}}{4p} +
\frac{1}{\rho_{c}}\big(\frac{1}{2}y^{2}+V(\phi)\big)\Big) \times \\
&\times&  \Big(\cos^{2}\big(\sqrt{\frac{\Delta}{p}}\frac{l_{0}}{2}\big) +
(1+\gamma^{2})\frac{\Delta l_{0}^{2}}{4p} -
\frac{1}{\rho_{c}}\big(\frac{1}{2}y^{2}+V(\phi)\big)\Big) \\ 
 &-& \frac{1}{2}\Big(\frac{1}{2}\sqrt{\frac{\Delta}{p}}l_{0}
\sin{(\sqrt{\frac{\Delta}{p}}l_{0})} - (1+\gamma^{2})\frac{\Delta l_{0}^{2}}{2p}
+ \frac{3}{\rho_{c}}y^{2}\Big) \times \\ 
 &\times& \Big(\cos{(\sqrt{\frac{\Delta}{p}}l_{0})}+
(1+\gamma^{2})\frac{\Delta
l_{0}^{2}}{2p} - \frac{2}{\rho_{c}}(\frac{1}{2}y^{2}+V(\phi))\Big)\Bigg\}
\end{eqnarray}
together with the constrain
\begin{equation}
\frac{1}{4}z^{2} =
\frac{\kappa}{3}\rho_{c}p^{2}\Big(\sin^{2}\big(\sqrt{\frac{\Delta}{p}}\frac{l_{0}}{2}\big)
-(1+\gamma^{2})\frac{\Delta l_{0}^{2}}{4p} +
\frac{\rho_{\phi}}{\rho_{c}}\Big) 
\Big(\cos^{2}\big(\sqrt{\frac{\Delta}{p}}\frac{l_{0}}{2}\big) +
(1+\gamma^{2})\frac{\Delta l_{0}^{2}}{4p} -
\frac{\rho_{\phi}}{\rho_{c}}\Big).
\end{equation}
The condition of the positivity of this expression implies that for a constant
value of $p$ the motion of the dynamical system in the phase plane
$(\phi,\dot{\phi})$ is restricted to
\begin{equation}
\rho_{c}\Big((1+\gamma^{2})\frac{\Delta l_{0}^{2}}{4p} +
\cos^{2}\big(\sqrt{\frac{\Delta}{p}}\frac{l_{0}}{2}\big)\Big) \geq
\frac{1}{2}y^{2} + \frac{1}{2}m^{2}\phi^{2} \geq
\rho_{c}\Big((1+\gamma^{2})\frac{\Delta l_{0}^{2}}{4p} -
\sin^{2}\big(\sqrt{\frac{\Delta}{p}}\frac{l_{0}}{2}\big)\Big)
\end{equation}
In Fig. \ref{phiphidot} this region is represented by the shaded field (in this example
$p=10$). For a given time region allowed for the motion is in the form of a ring. External
edge is due to quantum gravity effects while internal one coming from curvature. Radius of
internal edge decreases for increasing $p$. In Fig. \ref{phiphidot} we show exemplary
trajectories on the $(\phi,\dot{\phi})$ phase portrait. Two general types of trajectories 
are shown there. First type of trajectories is initialised in the point of maximal displacement
of the scalar field in the potential, then $\dot{\phi}=0$. Then filed falls down through the 
intermediate slow-roll regime, leading to the phase of inflation. This phase of inflation is 
seen as nearly horizontal attractor line in Fig. \ref{phiphidot}. It can be seen also trajectories 
starting from lower displacements are directed towards this inflationary attractor. These 
trajectories are shown also in Fig. \ref{pphi} where corresponding evolution on the $(\phi,p)$
plane is presented. 

Trajectories of the second kind are initialised in the bottom of the potential, 
where $\phi=0$. These trajectories are also directed to the same inflationary attractor.
It is worth to analyse this behaviour in more details. Considered  evolution is 
represented by the curves starting at $\phi=0$ in Fig. \ref{pphi}.
Here initial state is given in the contracting pre-bounce universe. Field is initially in the
bottom of the potential well. In this stage, value of the Hubble factor is neglected with respect 
to $m^2$. Therefore field undergoes certain oscillations.  It can be seen that is this stage field 
behaves like a dust matter. Universe behaves therefore like contracting, matter dominated one. 
However when bounce is approached then $H$ grows and becomes significant for the dynamics of 
$\phi$. Since H is negative act as anti-friction term in equation of motion
\begin{equation}
\ddot{\phi}+3H\dot{\phi}+m^2\phi = 0
\end{equation}
driving $\phi$ up the potential well. For the considered trajectory, field is driven up to
$\phi \simeq \pm 1$ in Planck units.  Then typical slow-roll inflation takes place leading to 
exponential growth of $p$. Field, damped by the friction term, falls down the potential well 
and starts to oscillate finally. Then again matter-like stage occur. However in this point phase 
of reheating should stars. This phase however cannot be studied without extension of this model. 
Therefore physical analysis of dynamics should be stopped at this point.

\section{Remarks on inverse volume corrections}

In the previous sections we have considered models with quantum holonomy corrections. 
Here we are making some remarks on the possible modifications due to the so called
inverse volume corrections. In the flat models definition of these corrections 
is ambiguous since it deepens on the fiducial cell volume. However in the  
considered closed model, fiducial cell volume is fixed and inverse volume corrections
are well defined. Since effects of inverse volume corrections were extensively 
studied in literature we are not going to perform full analysis of them. 
Interested reader can find more information abut them in Ref. \cite{Vandersloot:2005kh,Ashtekar:2006es}.
In particular interesting feature of the inverse volume corrections is the 
phase of super-inflation \cite{Copeland:2007qt}. 

\begin{figure}[ht!]
\centering
$\begin{array}{cc}   
\includegraphics[width=7cm,angle=0]{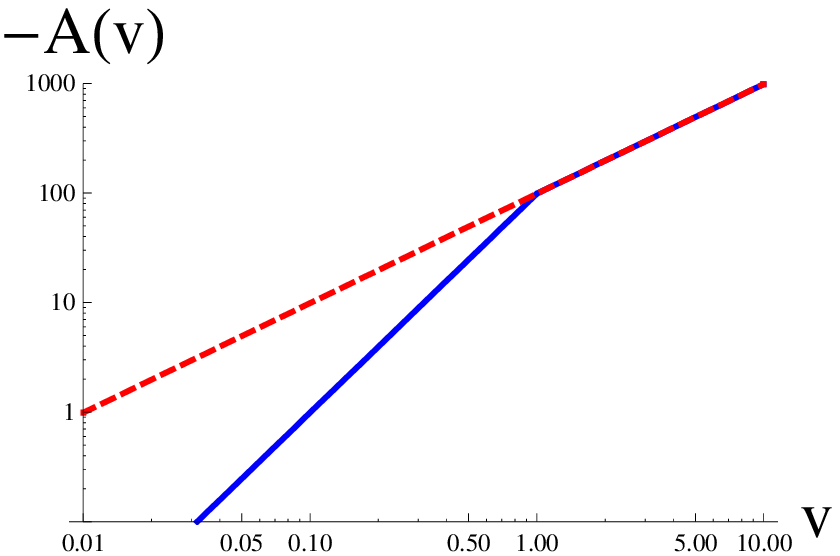}  &  \includegraphics[width=7cm,angle=0]{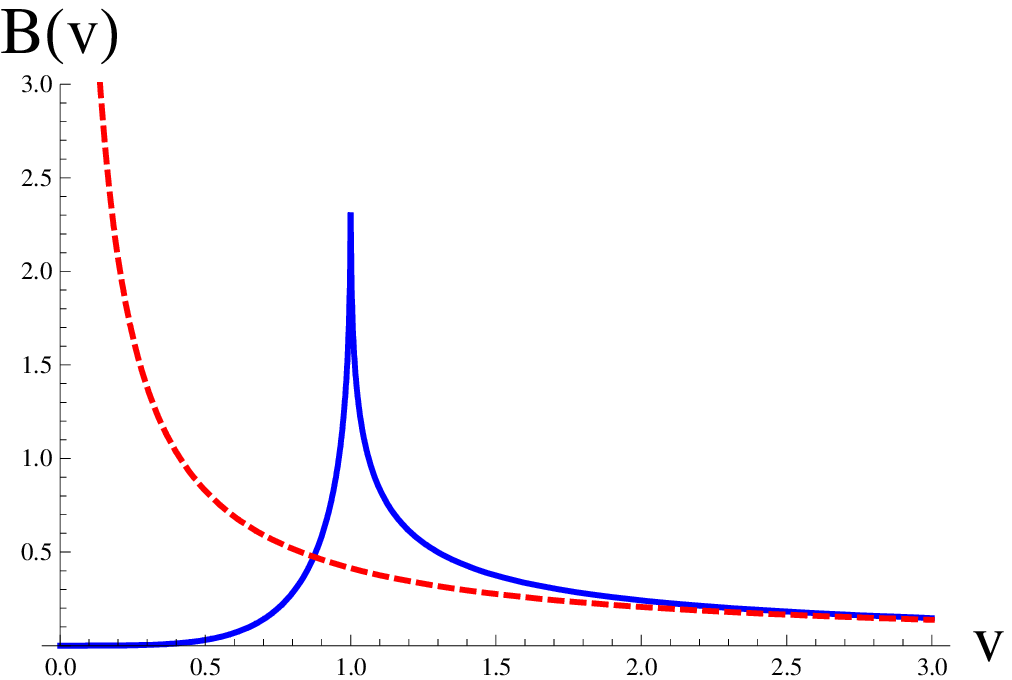}   
\end{array}
$
\caption{ 
{\bf Left }: Inverse volume correction $A(v)$ (straight line) together with 
approximation used in this paper (dashed line). 
{\bf Right }: Inverse volume correction $B(v)$ (straight line) together with 
approximation used in this paper (dashed line).}
\label{ABcorrections}
\end{figure}

Here we are going to show where effects of inverse volume corrections start to be important. 
We begin this analysis with general effective Hamiltonian derived in Ref. \cite{Ashtekar:2006es} 
\begin{eqnarray}
 \mathcal{H}_{\text{eff}} &=& \frac{A(v)}{16 \pi G} 
\left[  \sin^2\left(\bar{\mu} c - \bar{\mu}\frac{l_0}{2}  \right)   
- \sin^2 \left( \bar{\mu}\frac{l_0}{2}  \right)   +(1+\gamma^2) \frac{\bar{\mu}^2 l_0^2}{4}  \right] \nonumber \\
&+& \left( \frac{8\pi G \gamma}{6} \right)^{-3/2} B(v)\frac{p^2_{\phi}}{2}.    
\end{eqnarray}
Here functions $A(v)$ and $B(v)$ contains inverse volume corrections and are given by 
\begin{eqnarray}
A(v) &=& -\frac{27 K}{4\gamma^{3/2}} \sqrt{\frac{8\pi}{6}}  |v| \left| |v-1|-|v+1| \right|,  \label{Afull} \\
B(v) &=& \left(\frac{3}{2}\right)^3 K |v| \left| |v+1|^{1/3}-|v-1|^{1/3}\right|^3 \label{Bfull}
\end{eqnarray}
where 
\begin{equation}
K =  \frac{2\sqrt{2}}{3\sqrt{3\sqrt{3}}}.
\end{equation}
For $v \gg 1$ functions can be expanded as follows  
\begin{eqnarray}
A(v) &\simeq& -\frac{2\sqrt{48 \pi } v}{\gamma^{3/2} 3\sqrt{3}K}, \label{Aapprox} \\
B(v) &\simeq&  \frac{K}{v}. \label{Bapprox}
\end{eqnarray}
We show both functions (\ref{Afull}) and (\ref{Bfull}) with corresponding approximations 
(\ref{Aapprox}) and (\ref{Bapprox}) in Fig. \ref{ABcorrections}. Approximations (\ref{Aapprox}) 
and (\ref{Bapprox}) are precisely those applied in this paper, leading to initial Hamiltonian (\ref{Hamiltonian1}). 
Form Fig. \ref{ABcorrections} we see that corrections become important below $v\approx 2$.
Since relation between $p$ and $v$ variables is given by 
\begin{equation}
p(v)=\frac{\kappa  \gamma }{6}\left(\frac{v}{K}\right)^{2/3}
\end{equation}
we find $p(2)\simeq 2.8$. Therefore for $p\lesssim 3$ inverse volume corrections becomes significant.
So we have to be aware that results obtained in this paper are valid for $p\gtrsim 3$ and 
below this value other quantum effects should take place. However only small fraction of trajectories 
probe this region. In fact we can always choose sufficient initial conditions to avoid this region.
For instance for the model with a free scalar field, sufficiently high value of $p_{\phi}$ has to be 
chosen. Namely condition for $p > 3$ gives  $p_{\phi} > 18 \ l_{\text{Pl}}$. 

\section{Summary}

In the classical cosmology positive curvature introduces $-1/a^2$ term in the Friedmann 
equation. In the early universe other terms like radiation $1/a^4$ dominate and dynamical 
effect of curvature is negligible. However also quantum gravitational effects start 
to be important then and dynamics should be modified. Loop quantum cosmology gives us 
opportunity to study these effects. In particular in the effective formulation of LQC 
dynamics is governed by the classical equations with proper quantum corrections. These 
corrections modify also the curvature term  $-1/a^2$ and therefore dynamical 
significance of this term can be changed in the quantum regime. In this paper we have 
investigated whether curvature term could play significant role in early universe due 
to the quantum effects. We have found that curvature does not affect evolution in the 
quantum regime. In particular, maximal value of the Hubble factor is the same like in the 
flat case, namely $H_{\text{max}}=\sqrt{\frac{\kappa}{12}\rho_c}$.

In our considerations we concentrated on the models with cosmological constant, 
free scalar field and massive scalar field. In the models with cosmological constant
and free scalar field, dynamics reduce to 2D system. For the free field case, universe 
undergoes non-singular oscillations. This behaviour does not depend on the value of 
parameter $p_{\phi}$. However for the model with cosmological constant only,
type of solution depends on the value of $\Lambda$. For $\Lambda < \Lambda_{\text{c}}$ bouncing
solutions with asymptotic de Sitter  stages are obtained.  While  $\Lambda > \Lambda_{\text{c}} $
we find non-singular oscillations. Case $\Lambda = \Lambda_{\text{c}} $ is intermediate state when 
upper boundary reaches infinity. Here $\Lambda_c$ is critical value of $\Lambda$  and is given
by $\Lambda_{\text{c}} = \frac{\sqrt{3} m^2_{\text{Pl}}}{2\pi\gamma^3} \simeq 21 m^2_{\text{Pl}}$.
Emergence of this value is due to quantum modifications of the dynamics. Therefore observed 
change of the type of solution is of the quantum origin and does not appear in the classical case. 

Physically interesting consequences occur in the models with self-interacting scalar field.
Namely, the point is that, bounce gives mechanism to set good initial conditions for inflation. 
During the bounce field is driven up the potential well what is necessary to start phase of 
inflation. In the classical theory of inflation initial conditions have to be fixed suitably. 
Here they can be obtained from the different pre-bounce evolutions without need of any precise 
adjustment. This is nice feature of the bounce stage and is now confirmed for the flat and 
closed models in LQC.

\end{document}